\documentclass[pra,aps,twocolumn,showpacs]{revtex4}
\usepackage{graphicx}
\usepackage{amsfonts}
\usepackage{amssymb}
\usepackage{amsmath}
\usepackage{bm}
\usepackage{bbm}
\makeindex
\DeclareMathAlphabet\mathbfcal{OMS}{cmsy}{b}{n}

\newcommand{\onne}{\mathbbm{1}}
\newcommand{\bc}{\mathbf{c}}
\newcommand{\mk}{\mathbbm{k}}
\newcommand{\mM}{\mathbbm{M}}

\renewcommand{\r}{\textbf{r}}
\newcommand{\Y}{\textbf{Y}}
\newcommand{\E}{\textbf{E}}
\newcommand{\D}{\textbf{D}}
\renewcommand{\H}{\textbf{H}}

\renewcommand{\Re}{\mathrm{Re}}
\renewcommand{\Im}{\mathrm{Im}}

\newcommand{\one}{\hat{\bm 1}}

\newcommand{\grad}{\pmb{\nabla}}
\newcommand{\eps}{\hat{\pmb{\varepsilon}}}

\newcommand{\Eq}[1]{Eq.\,(\ref{#1})}

\newcommand{\kmax}{k_\mathrm{max}}
\newcommand{\kmaxS}{k_\mathrm{max}^\mathrm{S}}

\newcommand{\be}{\begin{equation}}
\newcommand{\ee}{\end{equation}}
\newcommand{\bea}{\begin{eqnarray}}
\newcommand{\eea}{\end{eqnarray}}

\newcommand{\Fig}[1]{Fig.\,\ref{#1}}
\newcommand{\Sec}[1]{Sec.\,\ref{#1}}
\newcommand{\App}[1]{Appendix\,\ref{#1}}
\newcommand{\Tab}[1]{Table\,\ref{#1}}
\newcommand{\Onlinecite}[1]{Ref.\,\onlinecite{#1}} 

\newcommand{\br}{\mathbf{r}}

\newcommand{\bE}{\mathbf{E}}
\newcommand{\bD}{\mathbf{D}}

\newcommand{\kapnu}{k}
\newcommand{\kapexactnu}{k^\mathrm{(exact)}}
\newcommand{\kapextranu}{k^{(\infty)}}

\newcommand{\Tdia}{T^{\mathrm{dia}}}
\newcommand{\Tinv}{T^{\mathrm{inv}}}

\begin{document}
\title{Resonant-state expansion applied to three-dimensional open optical systems:\\
A complete set of static modes}
\author{S.\,V. Lobanov}\email{LobanovS@cardiff.ac.uk}
\altaffiliation[Present address: ]{School of Medicine, Cardiff University, Cardiff CF24 4HQ, United Kingdom}
\affiliation{School of Physics and Astronomy, Cardiff University, Cardiff CF24 3AA, United Kingdom}
\author{W. Langbein}
\affiliation{School of Physics and Astronomy, Cardiff University, Cardiff CF24 3AA, United Kingdom}
\author{E.\,A. Muljarov}
\affiliation{School of Physics and Astronomy, Cardiff University, Cardiff CF24 3AA, United Kingdom}
\begin{abstract}

We present two alternative complete sets of static modes of a homogeneous dielectric sphere, for their use in the resonant-state expansion (RSE), a rigorous perturbative method in electrodynamics. Physically, these modes are needed to correctly describe the static electric field of a charge redistribution within the optical system due to a perturbation of the permittivity. We demonstrate the convergence of the RSE towards the exact result for a perturbation describing a size reduction of the basis sphere. We then revisit the quarter-sphere perturbation treated in [Doost {\it et al.}, Phys. Rev. A {\bf 90}, 013834 (2014)], where only a single static mode per each angular momentum was introduced, and show that using a complete set of static modes leads to a small, though non-negligible correction of the RSE result, improving the agreement with finite-element simulations. As another example of applying the RSE with a complete set of static modes, we calculate the resonant states of a dielectric cylinder, also comparing the result with a finite-element simulation.
\end{abstract}
\pacs{03.50.De, 42.25.-p, 03.65.Nk}
\date{\today}
\maketitle
\section{Introduction}
The resonant-state expansion (RSE) is a powerful theoretical method, recently developed in electrodynamics~\cite{MuljarovEPL10} for accurate calculation of resonant states (RSs) of an arbitrary open optical system. The concept of RSs presents a mathematically rigorous way of describing physical resonances of an open system, seen in its optical spectra, for example in the scattering cross-section. Using the Mittag-Leffler theorem, one can determine these spectra by expanding them into the RSs of the system \cite{LobanovPRA18}.
The RSs thus contain the full information about the system and ideally present a complete set of functions suited for expansion of any vector field within the volume of the optical system. This completeness of the RSs as well as the Mittag-Leffler expansion of the Green's diadic are at the heart of the RSE which allows one to accurately calculate the RSs of an optical system, using as basis the RSs of an unperturbed system which differs from the system of interest by a  perturbation of e.g. its permittivity~\cite{DoostPRA14}. To determine a complete set of RSs, it is advantageous to choose the basis system to be exactly solvable, such as a homogeneous dielectric sphere in vacuum, for which the analytical solutions in the form of Mie resonances are well known in the literature~\cite{LaiPRA90,DoostPRA14}.

When applying the RSE to three-dimensional (3D) open optical systems~\cite{DoostPRA14}, it has been recognized that in addition to the RSs $ \E_n(\r)$ describing physical resonances
and satisfying Maxwell's wave equations
\begin{equation}
\nabla\times \nabla\times\E_n = k_n^2\eps(\r)\E_n
\label{Eq:MWE}
\end{equation}
with outgoing boundary conditions,
one needs to include in the basis for the RSE also the zero-frequency modes $\bE^S_\lambda(\r)$, satisfying the static Maxwell's equation
\be
\nabla\times\bE^S_\lambda=0\,.
\label{curlE}
\ee
Here, index $n$ is used to label RSs, $\lambda$ to label static modes, $k_n=\omega_n/c$ is the RS wave number in vacuum, $\omega_n$ is the RS eigen frequency, $\eps(\r)$ is the permittivity tensor,
and the permeability is assumed to be $\mu=1$ for simplicity.
The electric fields $\bE^S_\lambda$ of the static modes do not satisfy the full set of Maxwell's equations, and therefore do not represent any physical states of the system. However, they are required for the Mittag-Leffler expansion of the Green's function, playing the role similar to that of the cut poles included for completeness into the RSE of 2D systems, for which the Green's function has cuts in the complex frequency plane~\cite{DoostPRA13}.  In the recent work~\cite{DoostPRA14}, a single static longitudinal electric mode  was added for each angular momentum. The added modes satisfy Maxwell's equation
\be \nabla\cdot\bD=0
\label{divD} \ee
both within and outside a dielectric sphere but violate Maxwell's boundary condition of the continuity of the normal component of the displacement $\bD$ across the sphere boundary. Adding only these modes to the basis for the RSE was suited to describe a homogeneous perturbation of the sphere permittivity, and only required for transversal magnetic modes which have a finite electric field normal to the interface. However, we found that adding only these static modes was insufficient to treat another simple perturbation of the dielectric sphere -- reducing its size.

In general, the RSs of any optical system satisfy \Eq{divD} at any point in space. This implies, in particular, that their electric fields have a non-zero divergence
\be
\nabla\cdot\bE=-\bE\cdot\frac{\nabla \varepsilon}{\varepsilon}\neq 0
\ee
in the regions of inhomogeneity of the permittivity $\varepsilon(\r)$. Obviously, this non-zero divergence cannot be reproduced by the RSE if all basis RSs respect
\be \nabla\cdot \E_n =0
\label{divE} \ee
inside the system. This is the case, however, for the RSs of a homogeneous sphere, which are known in the literature as transverse-electric (TE) and transverse-magnetic (TM) modes responsible for the resonant behavior in Mie scattering~\cite{Bohren1998}. Indeed, due to Maxwell's equation $\nabla \times\H_n=-ik_n\varepsilon \E_n$, \Eq{divE} holds everywhere except the sphere boundary. The static modes introduced in \Onlinecite{DoostPRA14} also have this property. One therefore needs additional modes in the basis with $\nabla\cdot\bE^S_\lambda\neq 0$, which can only be static modes, having $k^S_\lambda=0$.

In this paper, we present in \Sec{sec:Static-general} a general formulation of the problem of static modes of an open optical system and then introduce in \Sec{sec:Static-sphere} two sets of longitudinal electric static modes of a dielectric sphere in vacuum, both suited for a correct treatment of an arbitrary perturbation of its permittivity. One such set of modes is complementing the static modes already introduced in \Onlinecite{DoostPRA14}. Physically, these modes carry both volume charges within the sphere and surface charges on its boundary, so we call them volume-surface charge (VSC) modes. The other set introduces only volume charges, as they satisfy Maxwell's boundary condition for  $\D$ on the sphere boundary, and are therefore called volume-charge (VC) static modes. Note that for treating perturbations of the permeability $\mu$, similar sets of longitudinal static modes can be introduced for the magnetic field. In Secs.\,\ref{subsec:hom} and \ref{subsec:size}, we test both sets on the exactly solvable cases of, respectively, strength and size perturbations of a homogeneous sphere, also studying the convergence of the RSE towards the exact solution. We then revisit in \Sec{subsec:qsphere} the quarter-sphere perturbation and compare results with the previous calculation~\cite{DoostPRA14} and with finite-element simulations. Finally, in \Sec{subsec:cylinder} we use the RSE with static modes for calculating the RSs of a dielectric cylinder, also comparing results with a finite-element simulation.

\section{Static modes of an arbitrary open optical system
}\label{sec:Static-general}
We first consider an arbitrary finite open optical system in vacuum or in a homogeneous medium. Static modes $\E_\lambda^S$ of this system satisfy \Eq{curlE}, and therefore can be written as
\be
\E_\lambda^S(\r)=-\nabla\psi_\lambda(\r)\,.
\label{gradpsi}
\ee
Considering only square integrable solutions, the proper normalization of static modes, as it was derived in \Onlinecite{DoostPRA14}, takes the form
\be
\int \E_\lambda^S(\r)\cdot \D_\lambda^S(\r) d\r=1\,,
\label{norm}
\ee
where $\D_\lambda^S(\r) = \eps(\r)\E_\lambda^S(\r)$, and the integration is performed over the entire 3D space. Note that the factor of 2 difference compared to Eq.\,(7) of Ref.~\cite{DoostPRA14} was introduced in a generalized formulation of the RSE~\cite{MuljarovOL18}, which we use in the present work. Here we assume a frequency-independent permittivity; a generalization of \Eq{norm} for systems with frequency dispersion can be found in \Onlinecite{MuljarovOL18}. We note that a zero frequency pole in the  permittivity of a conductive material results in a vanishing amplitude of the normalized static modes and thus in their vanishing contribution.

Let us now consider an expression similar to \Eq{norm} for two static modes, $\lambda$ and $\lambda'$, including the case of the same mode $\lambda=\lambda'$,
\be
I_{\lambda\lambda'}=\int_V \E_\lambda^S\cdot \D_{\lambda'}^S d\r=
-\oint_{S_V} \psi_\lambda \D_{\lambda'}^S \cdot d{\bf S}+ \int_V  \psi_\lambda \nabla\cdot \D_{\lambda'}^S d\r\,,
\label{I}
\ee
where $V$ is an arbitrary (finite or infinite) volume which includes all the inhomogeneities of the optical system, and $S_V$ is its boundary. The right-hand side  of \Eq{I} is obtained by using \Eq{gradpsi} and the divergence theorem.
If we require, as boundary condition for the static modes, that $\psi_\lambda$ is vanishing on the surface $S_V$,
\be
\left. \psi_\lambda \right|_{S_V}=0\,,
\label{BC}
\ee
and outside it, for any state $\lambda$, then the surface term on the RHS of \Eq{I} vanishes, and the volume term can be seen as a scalar product which introduces a linear operator $\hat{L}$ such that
\be
\hat{L} \psi(\r)= -\nabla \cdot \eps(\r)\nabla \psi(\r) = \nabla\cdot \D(\r)=\rho(\r)\,.
\label{L}
\ee
The above quantity has the physical meaning of the free-charge density $\rho(\r)$ described by the electrostatic potential $\psi(\r)$.
Assuming these free charges can exist only within the system volume $V_0$ (included in $V$), results in the eigenvalue equation
\be
\hat{L} \psi_\lambda(\r)=\left\{ \begin{array}{cl}
\Lambda \psi_\lambda & {\rm within}\ V_0 \\
0 & {\rm otherwise}\,,
\end{array}\right.
\label{Lev}
\ee
where $\Lambda$ is the eigenvalue of the operator $\hat{L}$ corresponding to the eigenstate $\psi_\lambda$\,.
Equation~(\ref{Lev}) together with the boundary condition \Eq{BC} thus form a generalized Sturm-Liouville problem.
We note that by its physical meaning, the electrostatic potential $\psi_\lambda$ must be continuous across the system boundary. However, its spatial derivative does usually have a break across the system boundary which may lead to a presence of surface charges on the boundary, as discussed in detail in \Sec{sec:Static-sphere} below.

From \Eq{Lev} follows the orthonormality of static modes,
\bea
I_{\lambda\lambda'}&=&\int_{V}  \psi_\lambda \hat{L}\psi_{\lambda'} d\r\nonumber\\
&=&\Lambda\int_{V_0}\psi_\lambda \psi_{\lambda'} d\r=\Lambda'\int_{V_0}\psi_\lambda \psi_{\lambda'} d\r=\delta_{\lambda\lambda'}
\eea
where $\delta_{\lambda\lambda'}$ is the Kronecker delta, since $\Lambda\neq\Lambda'$ for different modes, or in case of degeneracy the modes can be made orthogonal by symmetry.
Furthermore, taking the same scalar product with the wave functions of the RSs, we see that static modes are orthogonal to all of the RSs of the optical system:
\be
\int_V \E_\lambda^S\cdot \D_n d\r=
-\oint_{S_V} \psi_\lambda \D_n \cdot d{\bf S}+ \int_V  \psi_\lambda \nabla\cdot \D_n d\r=0\,,
\label{ED}
\ee
due to the boundary condition \Eq{BC} for static modes and $\nabla\cdot \D_n =0$ for the RSs. Note that this argument is valid for a finite volume $V$ only. For $V\to\infty$, the integral \Eq{ED} may diverge. However, a proper orthogonality condition for this case, involving surface integrals, is provided in \Onlinecite{DoostPRA14}.

Finally, we note that in addition to the volume charge density of a static mode $\lambda$, given by
\be
\rho_\lambda(\r)=\Lambda\psi_\lambda(\r)\,,
\label{charge}
\ee
which is present only within the system volume $V_0$, there is a surface charge density
\be
\sigma_\lambda=-\left.\D_\lambda^S\right|_{S^-_V}
\ee
on the inner side of the boundary $S_V$, which is due to the fact that the normal component of the displacement $\D_\lambda$ is discontinuous across the boundary, owing to the boundary condition \Eq{BC}. The volume $V$ can be any, and if it coincides with the system volume $V_0$, there is a surface charge density on the system boundary, associated with each static mode. If instead $V\to\infty$, the surface charge density $\sigma_\lambda\to0$, due to $\D_\lambda^S$ vanishing quick enough at $\r\to\infty$. Below we consider these situations in detail for the analytically solvable case of a dielectric sphere.

\section{Static modes of a dielectric sphere in vacuum
}\label{sec:Static-sphere}

Consider a dielectric sphere in vacuum, having radius $R$ and permittivity $\epsilon$. The system is described by the permittivity tensor $\eps(\r)=\varepsilon(r) \one$, where
\be
\varepsilon(r)=(\epsilon-1)\Theta(R-r) +1
\label{perm}
\ee
and $\Theta(x)$ is the Heaviside step function.
Let us initially take the volume $V$ in \Eq{I} to be the infinite volume of the full 3D space.
Owing to the spherical symmetry of the system, we can make the ansatz
\be
\psi_\lambda(\r)=f_\lambda(r)Y_{lm}(\Omega)\,,
\label{psi-spher}
\ee
where $Y_{lm}(\Omega)$ are spherical harmonics (for definition see \App{App:VSH}). For a given fixed angular momentum $l$, the eigenvalue problem \Eq{Lev} then takes the form
\be
(\nabla^2_r+\lambda^2) f_\lambda(r)=0\ \ \ \ {\rm for}\ r<R
%
%
\label{Lev2}
\ee
and
\be
\nabla^2_r f_\lambda(r)=0\ \ \  {\rm for}\ r>R\,,
\label{Lev3}
\ee
where
\be
\nabla^2_r=\frac{d^2}{dr^2}+\frac{2}{r}\frac{d}{dr}-\frac{l(l+1)}{r^2}
\label{Laplacian}
\ee
and $l$ is the angular momentum.
Here, we have redefined for convenience the eigenvalue according to $\Lambda=\epsilon \lambda^2$.

Using the boundary conditions that the electrostatic potential $\psi_\lambda(\r)$ is continuous everywhere,  finite at $\r \to 0 $, and vanishing at $\r\to \infty$, we find the
solution of the radial equations (\ref{Lev2}) and (\ref{Lev3}) with the permittivity  \Eq{perm} in the following form:
\be
f_\lambda(r)=A_\lambda\times
\left\{ \begin{array}{ll}
j_l(\lambda r) & r<R \\
j_l(\lambda R) (R/r)^{l+1} & r>R\,,
\end{array}\right.
\label{f}
\ee
where $j_l(x)$ is the spherical Bessel function of order $l$ and $A_\lambda$ is the normalization constant.

To find the secular equation determining the eigenvalues $\lambda$ we require that the scalar product \Eq{I} between different static modes $\lambda\neq\lambda'$ is vanishing: $I_{\lambda\lambda'}=0$. For the potentials \Eq{psi-spher} we obtain
\bea
\E_\lambda^S(\r)&=&-\nabla\psi_\lambda(\r)=-f_\lambda(r)\nabla Y_{lm}(\Omega)- Y_{lm}(\Omega) \nabla f_\lambda(r)  \nonumber\\
&=& -\sqrt{l(l+1)}\frac{f_\lambda(r)}{r} \Y_{2lm}(\Omega)-\frac{d f_\lambda(r)}{dr} \Y_{3lm}(\Omega)\,,\nonumber\\
\label{Elam}
\eea
where $\Y_{ilm}(\Omega)$ are the vector spherical harmonics (for their definition and properties, see \Onlinecite{LobanovPRA18} and \App{App:VSH}). Using their orthonormality, we obtain
\bea
I_{\lambda\lambda'}&=&\int_0^\infty \varepsilon(r) r^2 dr \left[\frac{d f_\lambda}{dr}\frac{d f_{\lambda'}}{dr}+l(l+1)\frac{f_\lambda}{r}\frac{f_{\lambda'}}{r}\right]
\nonumber\\
&=& \frac{R}{\lambda^2-{\lambda'}^2} (F_\lambda G_{\lambda'}-F_{\lambda'} G_{\lambda})=0\,,
\label{sec}
\eea
where
\bea
F_\lambda&=&A_\lambda\lambda^2 j_l(\lambda R)\,, \label{Flam}\\
G_\lambda&=&A_\lambda[\epsilon\lambda R j'_l(\lambda R)+(l+1)j_l(\lambda R)]\,,
\label{Glam}
\eea
and $j'_l(x)$ is the derivative of the spherical Bessel function.

The secular equation (\ref{sec}) determines the eigenvalues $\lambda$ of the static modes.  Obviously, it is fulfilled if $F_\lambda=0$ or $G_{\lambda}=0$, which are the two special cases determining, respectively, the VSC and VC static modes, considered below in detail. A more general solution of \Eq{sec} is given by
\be
\alpha F_\lambda +\beta G_{\lambda}=0
\label{gen-case}
\ee
with arbitrary $\lambda$-independent constants $\alpha$ and $\beta$. The condition \Eq{gen-case} can also be written as
\be
\alpha \lambda^2 f_\lambda(R) -\beta R\left[\varepsilon(r)\frac{df_\lambda(r)}{dr}\right]^{R_+}_{R_-}=0\,,
\ee
where $R_\pm=R\pm0_+$ with a positive infinitesimal $0_+$.

Finally, the normalization constants $A_\lambda$ are found from the diagonal elements $I_{\lambda\lambda}$, defined by \Eq{I}, which have the following explicit form
\be
I_{\lambda\lambda}=\frac{R}{\lambda^2} F_\lambda G_{\lambda}+\epsilon\lambda^2 \int_0^R f_\lambda^2(r)r^2 dr =1\,,
\label{norm-gen}
\ee
with an analytical integral in the second term.

\subsection{Volume-charge (VC) static modes}

The set of VC static modes is generated by the condition
\be
G_{\lambda}=0
\ee
with $G_{\lambda}$ given by \Eq{Glam}, which
determines the eigenvalues $\lambda$. This condition physically implies that Maxwell's boundary condition of the continuity of the normal component of the displacement $\D_\lambda$  across the sphere boundary is fulfilled, meaning that there are no free surface charges carried by the static modes. This can be easily seen from \Eq{BC} used for $\alpha=0$ and $\beta=1$.  The (volume) charge density is given by
\Eq{charge}:
\be
\rho_\lambda(\r)=\epsilon \lambda^2 \psi_\lambda (\r) \Theta(R-r)\,,
\label{rhoVC}
\ee
where $\psi_\lambda (\r)$ is provided by Eqs.\,(\ref{psi-spher}) and (\ref{f}).
We therefore call this set of static modes the volume-charge (VC) basis.
The normalization condition \Eq{norm-gen} reduces to the second term only, and the normalization constants are given by
\be
A^2_\lambda=\frac{2}{\epsilon\lambda^2 R^3}\left[j_l^2(\lambda R)-j_{l-1}(\lambda R)j_{l+1}(\lambda R)\right]^{-1}\,.
\label{normVC}
\ee

\subsection{Volume-surface charge (VSC) static modes}
\label{Sec:VSC}

The set of VSC static modes is generated by the condition
\be
F_{\lambda}=0
\ee
with $F_{\lambda}$ given by \Eq{Flam}, determining another set
of eigenvalues $\lambda$. This condition, in turn, splits into two cases:
\be
j_l(\lambda R)=0\ \ \ \ {\rm and} \ \ \ \ \lambda=0\,.
\label{F2}
\ee
The first one determines an infinite set of modes with $\lambda\neq0$ found from zeros of the Bessel function $j_l(x)$, and the wave functions are given by
\be
\psi_\lambda(\r)=A_\lambda j_l(\lambda r) \Theta(R-r) Y_{lm}(\Omega)\,.
\ee
The second one is nothing else than the single static mode introduced in \Onlinecite{DoostPRA14}  for each $l$. Its wave function can be obtained by taking the limit $\lambda\to0$
in \Eq{f}, which gives
\be
f_0(r)=\tilde{A}_0\times
\left\{ \begin{array}{ll}
(r/R)^{l} & r<R \\
(R/r)^{l+1} & r>R\,.
\end{array}\right.
\label{f0}
\ee
Both type of modes, $\lambda=0$ and $\lambda\neq0$, violate Maxwell's boundary condition for $\D_\lambda$ at $r=R$ which implies the presence of a free surface charge for each static mode at the sphere boundary. The electric charge density in this case is given by
\be
\rho_\lambda(r)=\epsilon \lambda^2 \psi_\lambda (\r) \Theta(R-r)+ \epsilon \left.\frac{\partial \psi_\lambda (\r) }{\partial r}\right|_{R_-} \delta(R-r)\,,
\label{rho-full}
\ee
where the first term, having the same form as in \Eq{rhoVC}, corresponds to the volume charge, while the second term describes the surface charge.
We therefore call this set of modes the volume-surface charge (VSC) basis.

For $\lambda\neq0$ modes, the normalization is given by \Eq{normVC}, which can be simplified, using \Eq{F2}, to
\be
A^2_\lambda=-\frac{2}{\epsilon\lambda^2 R^3j_{l-1}(\lambda R)j_{l+1}(\lambda R)}\,.
\ee
The $\lambda=0$ mode normalization is instead given by
\be
\tilde{A}^2_0=\frac{1}{R(\epsilon l +l+1)}\,,
\ee
which is produced by the surface term in \Eq{norm-gen}, evaluated at $\lambda\to0$. Using these normalization constants allows us to obtain explicit expressions for the surface charge density
\be
\sigma_\lambda (\Omega)\equiv\epsilon \left.\frac{\partial \psi_\lambda (\r) }{\partial r}\right|_{R_-}
\ee
which appears in \Eq{rho-full}:
\bea
\sigma_\lambda (\Omega)&=&\sqrt{\frac{2\epsilon}{R^3}} Y_{lm}(\Omega)\ \ \ \ {\rm for}\ \ \ \lambda\neq0\,, \nonumber\\
\sigma_0 (\Omega)&=&\sqrt{\frac{\epsilon l +l+1}{R^3}} Y_{lm}(\Omega)\ \ \ \ {\rm for}\ \ \ \lambda=0\,. \nonumber
\eea
We see that all $\lambda\neq0$ modes are zero outside the volume $V_0$ of the dielectric sphere and produce volume and surface charges, while the $\lambda=0$  mode is non-zero in all space and produces only a surface charge on the sphere surface $S_{V_0}$.

As it was shown in \Onlinecite{DoostPRA14}, to use merely the $\lambda=0$ modes, having only the surface free charge as it is clear from \Eq{rho-full}, is sufficient for treating a homogeneous perturbation across the sphere with a step at $r=R$, which creates effective charges only at the sphere surface. Since this step is present in most applications using the homogeneous sphere as basis, it is expected to be more computationally efficient to use the VSC basis of static modes, as they provide a surface charge density on the surface of the basis sphere, which in the VC basis is harder to reproduce.

\section{Application to systems with scalar dielectric susceptibility}\label{sec:app}

In this section we consider the application of the RSE with both VS and VSC sets of static modes to various perturbations of a dielectric sphere. The perturbed system is described by a scalar permittivity,  $\eps(\br)+\Delta\eps(\br)=\hat{\mathbf{1}} [\varepsilon(\br)+\Delta\varepsilon(\br)]$, with $\Delta\varepsilon(\br)$ being the perturbation of the dielectric constant. As basis system we use a homogeneous dielectric sphere of radius $R$ and $\epsilon=4$.
We consider several types of perturbations, namely, a homogeneous increase of the permittivity of the sphere in \Sec{subsec:hom}, a size reduction of the sphere in \Sec{subsec:size}, a quarter-sphere perturbation in \Sec{subsec:qsphere}, and a deformation of the sphere into a cylinder in \Sec{subsec:cylinder}.

We use the standard formalism of the RSE, as described in~\cite{MuljarovEPL10,DoostPRA12,DoostPRA13,DoostPRA14,MuljarovOL18,LobanovPRA18}. Some details of the technical implementation and optimization of the inclusion of a large number of static modes are given in \App{App:RSE}.

\begin{figure}[t]
	\includegraphics*[width=\columnwidth]{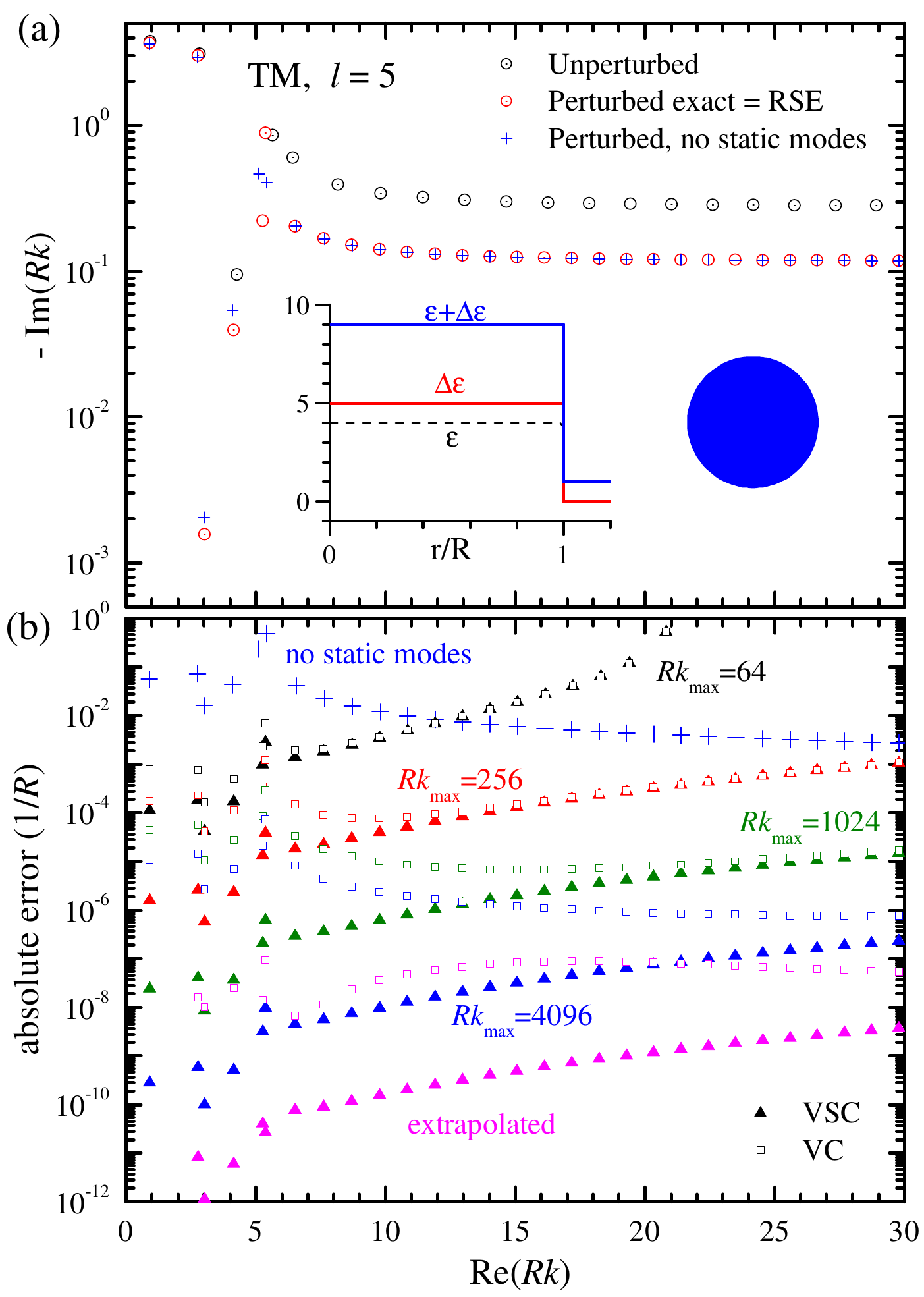}
	\caption{TM RSs with $l=5$ and a fixed $m$ for a homogeneous perturbation \Eq{eps-hom}, from the basis sphere with $\epsilon=4$ to the perturbed sphere with $\epsilon+\Delta\epsilon=9$. (a) Perturbed RS wave numbers calculated using the RSE with $R\kmax=4096$ without static modes (blue crosses), as well as using the exact secular equation or the full RSE (red circles with dots). The RS wave numbers of the basis system are shown as black circles with dots. Inset: Dielectric constant profile of the basis system (black dashed line), perturbed system (blue line), and the perturbation (red line). (b)  Error in the perturbed wave numbers calculated with no static modes (crosses), with VSC modes (triangles), and with VC modes (squares), for different $\kmax$ as labeled and color coded, as well as extrapolated using $R\kmax=4096$.} \label{fig:SphereHom}
\end{figure}

\subsection{Homogeneous sphere perturbation}
\label{subsec:hom} The perturbation we consider here is a homogeneous change of the permittivity over the whole sphere, given by
\be \Delta\varepsilon(\br)=\Delta\epsilon\,\Theta(R-r)
\label{eps-hom}
\ee
with the strength of $\Delta\epsilon=5$ used in the numerical calculation. This perturbation is spherically symmetric, so that RSs of different angular quantum numbers $(l,m)$, and different transverse polarization (TE or TM) do not mix, and are degenerate in $m$, making the RSE problem effectively 1D. We show in \Fig{fig:SphereHom}, for illustration, RSs with $l=5$, as in \Onlinecite{DoostPRA14}. Spherically symmetric perturbations do not couple static modes to TE RSs, since for the latter only the $\Y_{1lm}$ component is non-vanishing in the basis of the vector-spherical harmonics (see~\cite{LobanovPRA18} for details), while the static modes have the $\Y_{1lm}$ component vanishing, see \Eq{Elam}. We therefore show here only the TM RSs, which couple to the static modes by a spherically symmetric perturbation, as both types of fields have a non-vanishing radial component. For the VSC set, only the $\lambda=0$ mode couples to the TM RSs by the perturbation \Eq{eps-hom}. Indeed, the matrix elements between TM RSs and all other statics modes are proportional to the volume integral in \Eq{ED} and thus vanish. For the VC set instead, all static modes are coupled to the TM RSs. The perturbed TM RSs obey the same secular equation as the basis system (see Eq.(31) in \Onlinecite{DoostPRA14}) with the new permittivity of the sphere $\epsilon+\Delta\epsilon=9$, so that the new wave numbers $\kapnu$ calculated using the RSE can be compared with the exact values $\kapexactnu$ obtained from that secular equation.

Without any static modes, the error in the wave numbers $\kapnu$ of the perturbed RSs remains large, in the $10^{-1}$ to $10^{-2}$ range, see \Fig{fig:SphereHom}(b), consistent with the results shown in \Onlinecite{DoostPRA14}.
Using the VSC set, the error of $\kapnu$ is reduced to values below $10^{-7}$ for $R\kmax=4096$. We find that as we increase $N$, the error $\bigl|\kapnu-\kapexactnu\bigr|$ scales as $\kmax^{-3}$.  Figure~\ref{fig:SphereHom}(a) shows the resulting new wave numbers $\kapnu$ and \Fig{fig:SphereHom}(b) shows their errors for $R\kmax=$ 64, 256, 1025, 4096 (corresponding to $N_1=$ 40, 164, 652, 2608, respectively) and $R\kmaxS=$ 397, 1586, 6344, 25377 (corresponding to $N_2=$124, 503, 2017, 8076, respectively), see \App{App:RSE} for details. Note, however, that the only static mode relevant for this perturbation is $\lambda=0$, as explained above (but technically we include here for consistency the full VSC set).  Following the procedure described in \Onlinecite{DoostPRA12}, we extrapolate the new wave numbers to infinite $\kmax$ yielding $\kapextranu$.   We find that this extrapolation provides 1 to 2 orders of magnitude further reduction of the error, see \Fig{fig:SphereHom}(b).

Using instead the VC static modes (with $R\kmaxS$=430, 1718, 6873, 27492, corresponding to $N_2=$134, 544, 2185, 8748, respectively),  the error of $\kapnu$, especially for RSs with small wave numbers (see \Fig{fig:SphereHom}(b)) is reduced to only $10^{-4}$. Again, the extrapolation to $\kapextranu$ provides 1 to 2 orders of magnitude reduction of the error.
We find that the error for small $|R\kapnu|$, which is dominated by the static modes, scales as $\kmax^{-1}$, much slower than with the VSC basis. This is a consequence of the slow convergence in description of the effect of a surface charge induced by the perturbation (represented by a delta function $\delta(R-r)$)  when no surface charges are present in the basis, which is the case of the VC set.

\begin{figure}[t]
	\includegraphics*[width=\columnwidth]{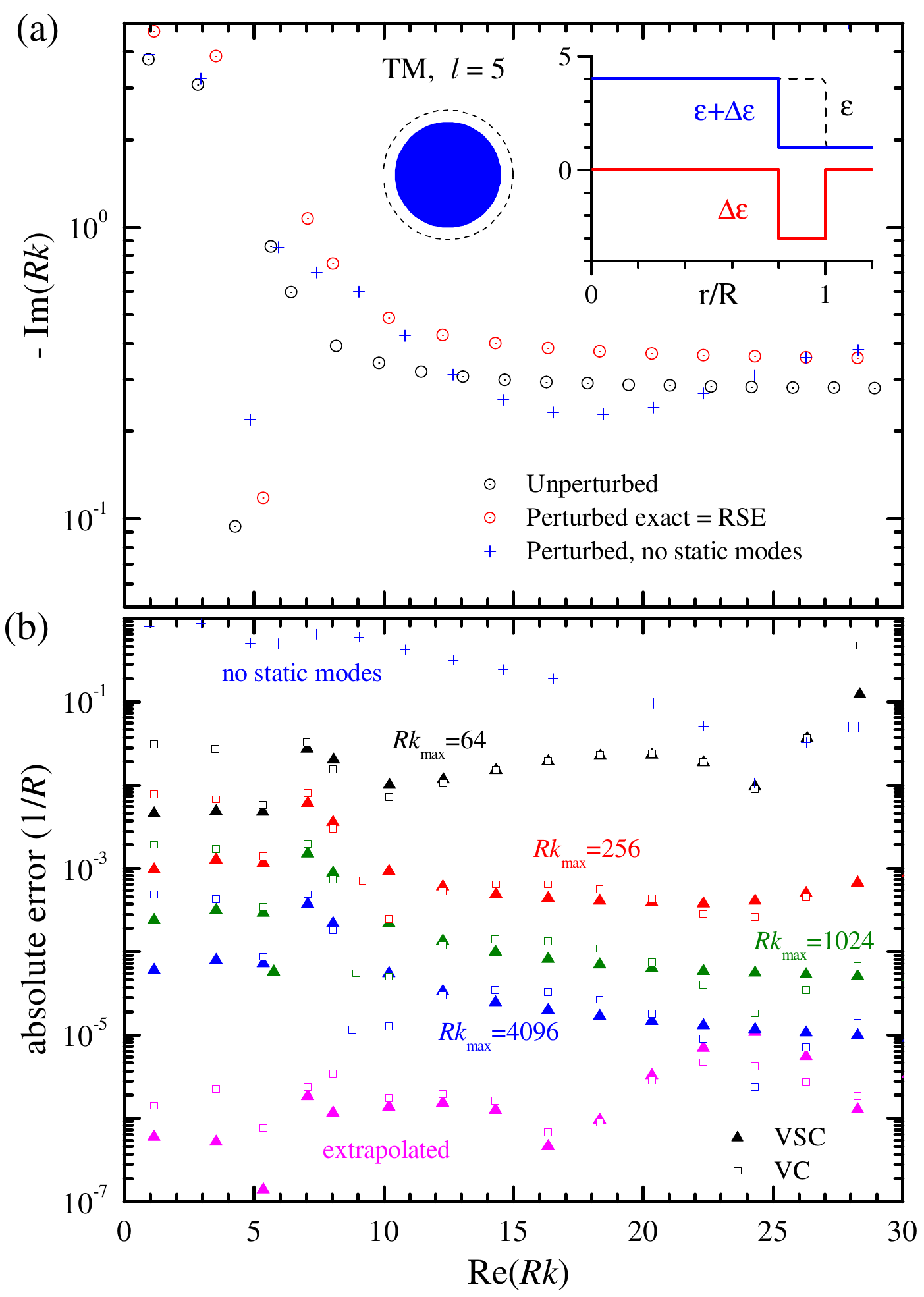}
	\caption{As Fig.\,1 but for the size perturbation \Eq{eps-size} to a sphere with the radius of $0.8\,R$.
}
\label{fig:SphereSize}
\end{figure}

\subsection{Size perturbation of a sphere}
\label{subsec:size}

We now consider a perturbation amounting to a size reduction of the sphere by 20\%, given by
\be \Delta\varepsilon(\br)=(1-\epsilon)\Theta(R-r)\Theta(r-0.8R)\,.
\label{eps-size}
\ee
This perturbation is also spherically symmetric, and we again show $l=5$ TM RSs in \Fig{fig:SphereSize}.  The perturbed RSs obey the same secular equation for a sphere (Eq.(31) in \Onlinecite{DoostPRA14}), with the radius reduced to $0.8R$, so that the new wave numbers $\kapnu$ calculated using the RSE can be again compared with the exact values $\kapexactnu$. The RSE wave numbers for $R\kmax=4096$ are shown in \Fig{fig:SphereSize}(a). Without static modes, the error stays large, above $10^{-2}$, similar to the homogeneous perturbation. Adding either the VSC or the VC set, the error of $R\kapnu$ is reduced to below $10^{-4}$, and the extrapolation to $\kapextranu$ provides 1 to 2 orders of magnitude further reduction of the error. We find that for large $\kmax$, the relative error scales as $(\kmaxS)^{-1}$ in both cases. This is due to the step of the perturbation at $r=0.8R$, which leads to a surface charge inside the basis sphere which is not provided by a single state in both sets. Notably, the error using the VC set shows oscillations versus Re$(Rk)$, which are absent for the VSC set. This is attributed to the contributions of the two surface charges at $r=R$ and $r=0.8R$, which lead to an interference in the matrix elements in the VC set. For the VSC set instead, the surface charge at $r=R$ is coupling only to the $\lambda=0$ mode, and the surface charge at $r=0.8R$ is coupling only to the $\lambda\neq0$ modes, so that no interference is present.

\subsection{Quarter-sphere perturbation}
\label{subsec:qsphere}

\begin{figure}
	\includegraphics*[width=\columnwidth]{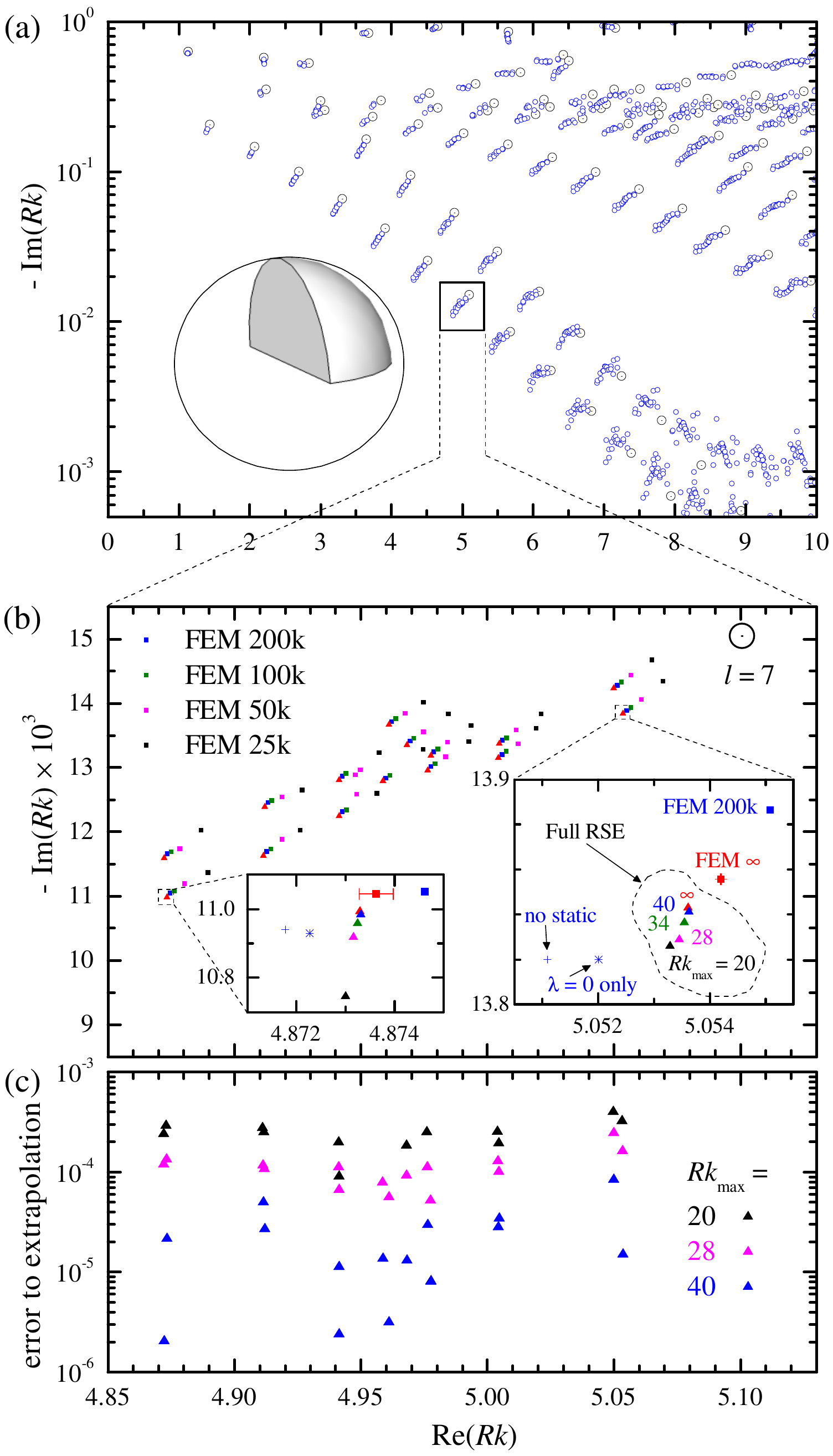}
	\caption{(a) Wave numbers of the basis RSs (black circles with dots) and the perturbed RSs (blue circles) for a quarter-sphere perturbation given by \Eq{eps-Quarter} with $\Delta\epsilon=1$, calculated by the RSE with $R\kmax=40$ using VSC static modes. A sketch of the structure is shown.  (b) Zoom of (a) showing the splitting due to the perturbation of $2l+1$ degenerate basis WGMs with $l=7$.
	The results of FEM simulations using 200k, 100k, 50k and 25k finite elements are shown for comparison. The insets show further zooms around individual RSs, containing additionally data from extrapolated FEM, extrapolated RSE, RSE using only the $\lambda=0$ static modes, and no static modes, as labeled. (c) Error of the RSE wave numbers using the extrapolated RSE as reference, as function of $\Re(Rk)$, calculated by the RSE with different $R\kmax$ as labeled, for the RSs shown in (b). }\label{fig:QSphere}
\end{figure}
We now revisit the example of the quarter-sphere perturbation treated in \Onlinecite{DoostPRA14}. This perturbation breaks the continuous rotational symmetry for both spherical angles $\theta$ and $\varphi$ and is thus not reducible to an effective one- or two-dimensional system. The perturbation is given by
\be \Delta\varepsilon(\br) = \Delta\epsilon\Theta(R-r)\Theta\left(\frac{\pi}{2}-\theta\right)\Theta\left(\frac{\pi}{2}-|\varphi|\right)
\label{eps-Quarter}
\ee
and corresponds physically to a uniform increase of the dielectric constant in a quarter-sphere volume, as sketched in \Fig{fig:QSphere}. For the results shown here, we use $\Delta\epsilon=1$, as in \Onlinecite{DoostPRA14}.  The calculation of the matrix elements requires numerical integration, as detailed in \Onlinecite{DoostPRA14}. The perturbation mixes RSs of different $l$, $m$, and polarization. The remaining mirror symmetry $\varphi \rightarrow -\varphi$ of the system decouples $m\geqslant0$ TE and $m<0$ TM RSs, having fields of odd parity ($-$), from $m<0$ TE and $m\geqslant0$ TM RSs, having fields of even parity ($+$). For this calculation we use the VSC basis, split according to the parity of the fields, with the same selection rules as TM RSs.

The lifting of the $m$-degeneracy of the new RSs provides a splitting of resonances in \Fig{fig:QSphere}(a) and (b). An analytic solution for this perturbation is not available, so that we estimate the error using as exact solution the extrapolated values of the largest $\kmax$. A convergence with a power law around $(\kmax)^{-3}$ is observed, resulting in relative errors in the $10^{-4}$ to $10^{-5}$ range.
The irregular arrangement of the perturbed whispering-gallery modes (WGMs) in the right-bottom part of \Fig{fig:QSphere}(a) is due to this remaining small error which manifests itself in fluctuating $\Im\,k$ well seen in the log scale used.

Exactly the same system was treated in \Onlinecite{DoostPRA14}, using only the $\lambda=0$ static modes, and was compared with FEM calculations using the commercial solver COMSOL, see Fig.\,4 of \Onlinecite{DoostPRA14}. These COMSOL results are also shown in the present \Fig{fig:QSphere} with $N_{\rm G}=25$k, 50k, 100k and 200k finite elements. With increasing $N_{\rm G}$, the COMSOL wave numbers converge, with an error scaling approximately as $N_{\rm G}^{-1}$. However, it was found in \Onlinecite{DoostPRA14} that increasing the precision of the RSE and COMSOL, a relative difference remained in the few times $10^{-4}$ range. One possible origin of this systematic deviation could be due to the incomplete static-mode basis used in \Onlinecite{DoostPRA14}. We have therefore repeated the RSE calculation using the VSC basis, as shown here. We have extrapolated both the COMSOL and the RSE results, as shown in the insets of \Fig{fig:QSphere}(b). We find that the remaining relative difference between the extrapolated COMSOL and RSE results is in the $10^{-4}$ range, which is still somewhat larger that the estimated error of the extrapolations shown in \Fig{fig:QSphere}(c). Nevertheless, the deviation between FEM and RSE results reported in \Onlinecite{DoostPRA14} was partly due to the incomplete basis used in that work. To explicitly identify the role of the static modes, we show in the insets of \Fig{fig:QSphere}(b) also the RSE results using no static mode, or using only the $\lambda=0$ static modes as in \Onlinecite{DoostPRA14}. We see that the full effect of the static modes on the perturbed RSs shown is about $5\times10^{-4}$ relative change.

\subsection{Sphere to cylinder perturbation}
\label{subsec:cylinder}

\begin{figure*}
	\includegraphics*[width=\textwidth]{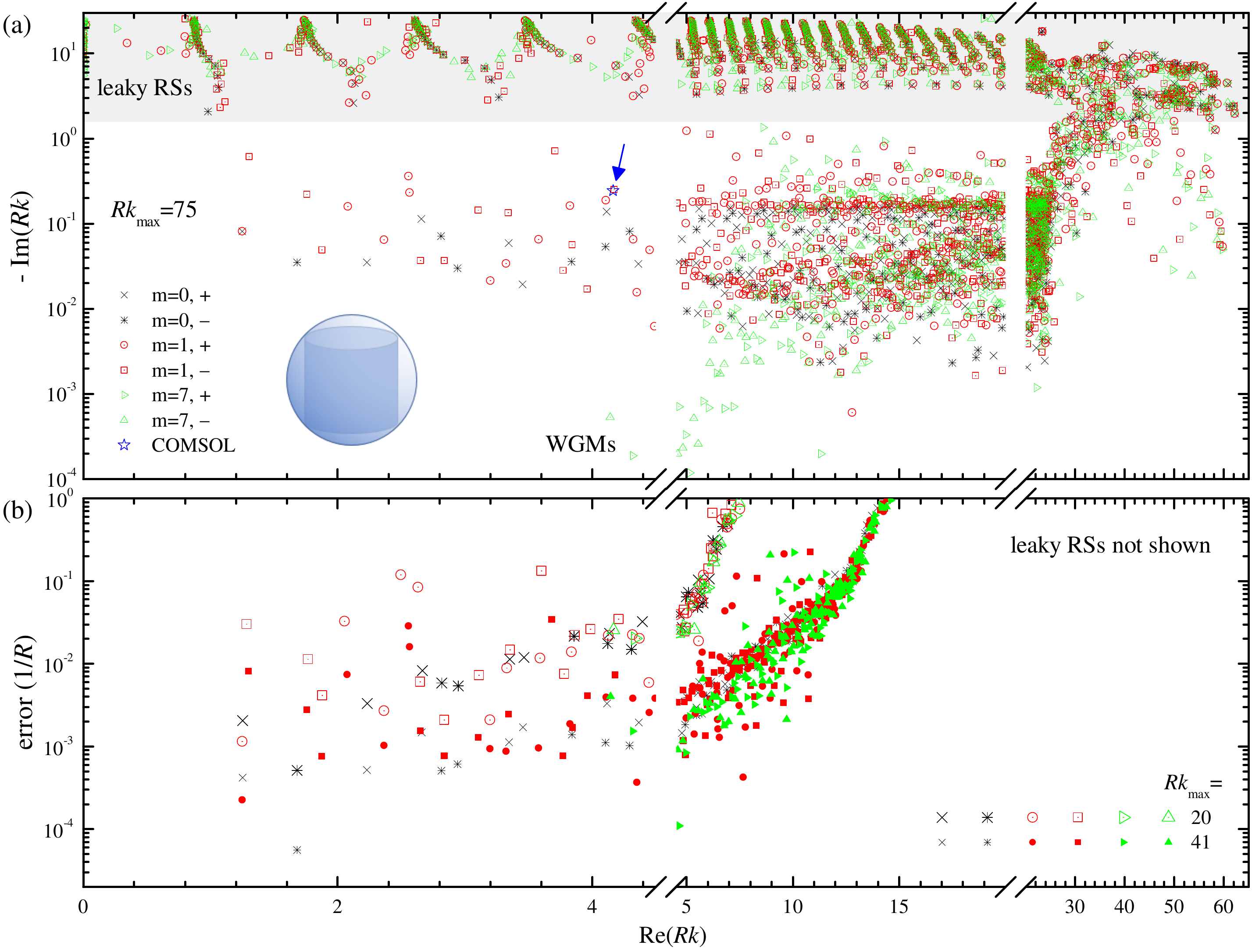}
	\caption{(a) RS wave numbers of a dielectric cylinder with radius $a$ and height $2a$, with azimuthal quantum numbers $m=0,1$, and 7, and parity $+$ and $-$ as labeled, calculated using the RSE with $R\kmax=75$ ($R=a\sqrt{2}$). The leaky RS having $-\Im(Rk)>1.5$ are separated by a grey shadow. (b) Error of the RS wave numbers for $R\kmax=20$ and 41, using the values for $R\kmax=75$ as reference, not showing results for the leaky modes.
} \label{fig:CylRS}
\end{figure*}

We consider here a perturbation which transforms a sphere into a cylinder of a height $2h$ equal to its diameter $2a$. It retains axial symmetry and inversion symmetry, and thus represents an effective 2D system. The perturbation is given by
\be \Delta\varepsilon(\br) = \Delta\epsilon\Theta(R-r)\left[\Theta\left(r|\cos\theta|-h\right)+\Theta(r\sin \theta-a)\right]\,,
\label{eps-cyl}
\ee
with $R=\sqrt{h^2+a^2}$ being the radius of the basis sphere. Owing to the axial symmetry and the mirror symmetry, the RSs of the cylinder have well defined $m$, and parity ($+,-$) of the field under the $z \rightarrow -z$ mirror imaging. The full perturbation matrix of the RSE is thus separated into blocks corresponding to these quantum numbers, significantly reducing the size of the eigenvalue problem.
Also note that the RSs of $m$ and $-m$ are degenerate due to the mirror symmetry $\varphi\rightarrow-\varphi$.

The basis sphere of radius $R$ has the same permittivity and just encloses the cylinder. To calculate the values of the symmetry-allowed matrix elements of the perturbation, we discretize the integrals into shell-segments. Shells of radial thickness $R/s$ are used, where $s$ is the number of shells. The polar angular ranges for each shell are then determined using the intersection of the center radius of the shell with the cylinder surface.
The number of shells is chosen using the Nyquist criterion of sampling, with 2 shells per period of $\kmax$, $s=\sqrt{\epsilon}\kmax R/\pi$. This choice ensures that with increasing $\kmax$, the spatial sampling of the structure is refined according to the spatial resolution of the basis RSs.

\begin{figure}
	\includegraphics*[width=\columnwidth]{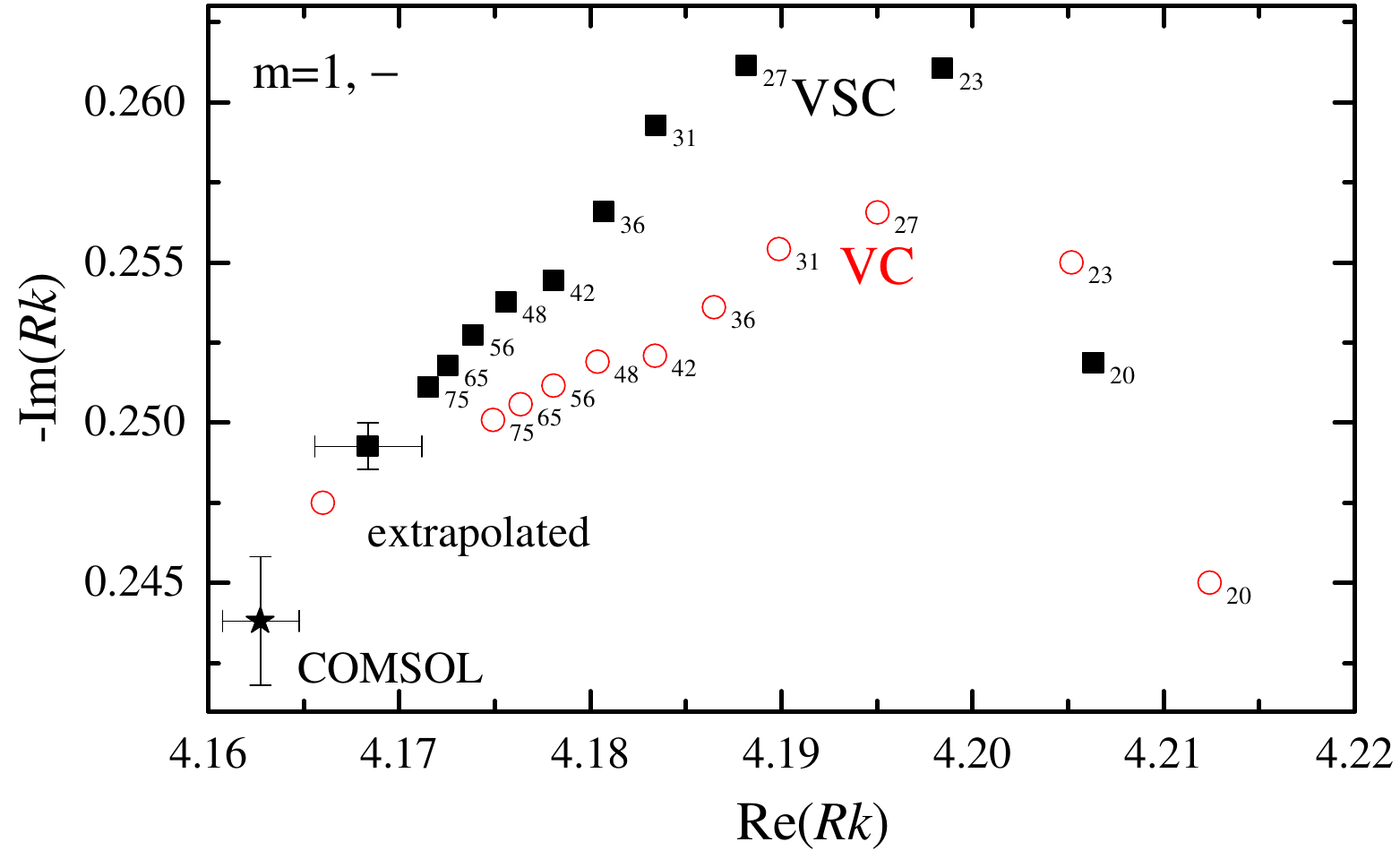}
	\caption{Convergence of the wave number for a selected RS of a cylinder, with $m=1$ and parity $-$ . Results by the RSE with different $R\kmax$ as labeled (close to the symbol),   both for the VSC (black squares) and VC (red open circles) set of static modes. The extrapolated values for the RSE and the value calculated using COMSOL~\cite{MuljarovPRB16} are also shown, along with their estimated errors.}\label{fig:CylKErr}
\end{figure}

The resulting RSs with $m=0,1$, and 7, and parity $+$ and $-$ are displayed in \Fig{fig:CylRS}(a) for the largest basis size considered, $R\kmax=75$ and $R\kmaxS=258$. The RSs with large losses, $-\Im(Rk)>1.5$, are called leaky modes. For the other RSs, we see that for small $m$-values ($m=0$ and 1), the losses $-\Im(Rk)$ are in the 0.1 range, increasing with $k$ their spread towards smaller losses and  reaching Q-factors above 1000, where $Q=-\Re\,k/(2\Im\,k)$. The low-loss RSs for these small values of $m$  can be formed at large $k$ by a field localized away from the edges and experiencing total internal reflection at the surfaces (center top and bottom, and center of the cylinder side wall). For the larger $m=7$ instead, we find that the lowest-frequency RSs have the character of WGMs, with $-\Im(Rk)<10^{-3}$ and Q-factors around $10^{4}$.
The error of the wave numbers, defined as the absolute difference to the values obtained for $R\kmax=75$, are given in \Fig{fig:CylRS}(b), not including the leaky modes. We find that RSs with $|Rk|\lesssim 5$ have the errors in the $10^{-2}$ range for $R\kmax=20$, and in the $10^{-3}$ range for $R\kmax=41$, thus scaling approximately as $\kmax^{-3}$, similar to the example with the quarter-sphere perturbation discussed above. Notably, as known about the RSE \cite{MuljarovEPL10}, for a given $\kmax$, the error is increasing with increasing wave number, limiting the range of small error to RSs with $|k|\lesssim\kmax/4$. We can see that within this range, the RSE is determining a large number of RS simultaneously, a few hundred for $R\kmax=75$.

To verify the wave numbers obtained for the cylinder using the RSE with static modes, we concentrate on the RS with $m=1$ and odd parity which has been calculated in \Onlinecite{MuljarovPRB16} using COMSOL, with a resonance frequency of $Rk= 4.16275 - 0.24382i$ and an estimated error of $0.002$ in real and imaginary part. We show in \Fig{fig:CylKErr} the wave number of this RS, and compare it with the results of the RSE using the VSC or the VC basis, as functions of $\kmax$, as well as with their extrapolated values including errors (see \Onlinecite{DoostPRA12} for the extrapolation procedure). We find that the convergence is somewhat different for the VC and VSC static basis, but the extrapolated values are equal within the estimated errors. The COMSOL result is in agreement with the extrapolated RSE within the estimated error, indicating that including the used sets of static modes indeed provides a complete basis for general 3D geometries.

\section{Summary}
In summary, we have introduced a general procedure for determining a full set of static modes for an arbitrary open optical system, supplementing the set of RSs, in order to form a complete basis for the RSE. Including this set is required to achieve high accuracy in the RSE in most cases, and specifically for general three-dimensional perturbations. We have shown in particular that static modes are required to treat perturbations which have a gradient of the permittivity in the direction of the RS electric field, thus creating effective free charges which are incompatible with Maxwell's boundary conditions of the basis system, and therefore not described by the RSs alone. Two alternative sets of static modes have been introduced for a dielectric sphere in vacuum, with one of them including free charges on the surface of the sphere, and thus being somewhat advantageous when treating perturbations with a permittivity step at the sphere surface. Using this complete basis in the RSE was then shown to result in convergence towards the analytical RSs of a reduced-size sphere, and towards the numerically determined RSs of a cylinder and a sphere with a quarter-sphere perturbation. Using the RS basis extended by including either of the two static mode sets, the RSE is expected to be numerically exact for general three-dimensional confined geometries.

\begin{acknowledgments}
This work was supported by the EPSRC under grant EP/M020479/1.
S. L. and E. A. M. acknowledge support from RFBR (Grant No. 16-29-03283).
\end{acknowledgments}

\appendix

\section{Scalar and vector spherical harmonics}
\label{App:VSH}

Following~\cite{DoostPRA14}, we define the scalar spherical harmonics $Y_{lm}(\Omega)$ as {\it real} functions:
\begin{equation}
Y_{lm}(\Omega) = \sqrt{\frac{2l+1}{2}\frac{(l-|m|)!}{(l+|m|)!}}P^{|m|}_l(\cos\theta)\chi_m(\varphi)\,,
\label{Eq:Y}
\end{equation}
where $\Omega=(\theta, \varphi)$ in the spherical coordinate system, $P^m_l(x)$ are the associated Legendre polynomials, $l$ and $m$ are the spherical quantum numbers, and
\be \chi_m(\varphi)=\left\{
\begin{array}{lll}
\pi^{-1/2}\sin(m\varphi) & {\rm for} & m<0\, \\
(2\pi)^{-1/2} & {\rm for} & m=0\, \\
\pi^{-1/2}\cos(m\varphi) & {\rm for} & m>0\,.
\end{array}
\right. \label{chi-n} \ee
This is done in order to satisfy the orthonormality condition without using the complex conjugate:
\be
\int Y_{lm}(\Omega)Y_{l'm'}(\Omega)d \Omega=\delta_{ll'}\delta_{mm'}\,,
\ee
where $d\Omega=\sin\theta d\theta d\varphi$.

The vector spherical harmonics  are defined following~\cite{LobanovPRA18} as
\bea
\Y_{1lm}(\Omega) &=& \frac{1}{\sqrt{l(l+1)}} \r \times \grad Y_{lm}(\Omega)\,, \label{Eq:Y1}\\
\Y_{2lm}(\Omega) &=& \frac{1}{\sqrt{l(l+1)}} r \grad Y_{lm}(\Omega)\,, \label{Eq:Y2}\\
\Y_{3lm}(\Omega) &=& \frac{\r}{r} Y_{lm}(\Omega)\,.\label{Eq:Y3}
\eea
They satisfy the following orthonormality condition:
\begin{equation}
\int \Y_{ilm}(\Omega)\cdot \Y_{i'l'm'}(\Omega)d \Omega=\delta_{ii'}\delta_{ll'}\delta_{mm'}\,.
\end{equation}

\section{RSE with static modes}
\label{App:RSE}

In \Onlinecite{DoostPRA14}, the RSE equation given by Eq.(13) therein was used to take into account the LE ($\lambda=0$) static mode, using a non-zero wave vector $Rk=10^{-7}$, since it was numerically more efficient than solving the generalized eigenvalue problem given by Eq.(12) therein. As we are now dealing with a large number of static RS, we instead follow a different approach, exploiting the zero wave vector of all static modes. We start with the generalized eigenvalue problem, given by Eq.(12) of \Onlinecite{DoostPRA14}, written in matrix form as~\cite{LobanovPRA18}
\begin{equation}
\mathbbm{k} \bc = \varkappa \mathbbm{M} \bc\,,
\label{mequ}
\end{equation}
where $\mk$ is a diagonal matrix containing the wave numbers of the basis RSs and static modes, $\mathbbm{M}=\onne+\mathbbm{V}$ with $\onne$ being the unit matrix and $\mathbbm{V}$ the perturbation matrix. $\varkappa$ and $\bc$ are, respectively, the eigenvalue and the eigenvector of a perturbed RS. The matrix elements of the perturbation matrix $\mathbbm{V}$ are given by
\begin{equation}
(\mathbbm{V})_{ij} = \int_{V_0} \E_{i}(\r) \cdot \Delta\eps(\r) \E_{j}(\r)d\r \,,
\label{Eq:Vnm}
\end{equation}
with $i$ and $j$ labeling both the RSs and the static modes of the unperturbed system. The RSs are normalized in accordance with~\cite{MuljarovOL18}, with the already mentioned factor of 2 difference compared to the earlier used normalization.

We now separate the notation explicitly into the $N_1$ RSs and $N_2$ static modes, so that the matrices $\mk$ and $\mM$ split into four sub-matrices, with $N_1$-dimensional square top-left sub-matrices $\mk_1$ and $\mM_{11}$ corresponding to the RSs. The eigenvector $\bc$ also splits into two sub-vectors, with an $N_1$-dimensional top sub-vector $\bc_{1}$ corresponding to the RSs and $N_2$-dimensional bottom sub-vector $\bc_{2}$ corresponding  to the static modes. The matrix equation (\ref{mequ}) then reads
\begin{equation}
\begin{pmatrix}
\mk_1 & 0\\
0 & 0
\end{pmatrix}
\begin{pmatrix}
\bc_{1}\\
\bc_{2}
\end{pmatrix}
= \varkappa
\begin{pmatrix}
\mM_{11} & \mM_{12}\\
\mM_{21} & \mM_{22}
\end{pmatrix}
\begin{pmatrix}
\bc_{1}\\
\bc_{2}
\end{pmatrix},\label{Eq:LargeEig}
\end{equation}
where $\mM_{12}$ is the transpose of $\mM_{21}$.  The above matrix equation then splits into a pair of matrix equations,
\bea
\mk_1\bc_{1}&=&\varkappa(\mM_{11}\bc_{1}+ \mM_{12}\bc_{2})\,,\\
0&=&\mM_{21}\bc_{1}+ \mM_{22}\bc_{2}\,,
\eea
which can be written as an $N_1 \times N_1$  eigenvalue problem for the RSs,
\be
\mk_1\bc_{1}= \varkappa \widetilde{\mM}_{11}\bc_{1}
\label{Eq:Diag}
\ee
with the effective perturbation matrix
\be
\widetilde{\mM}_{11}=\mM_{11}- \mM_{12}  \mM_{22}^{-1} \mM_{21}\,,
\ee
and an auxiliary equation for the static mode amplitudes
\be
\bc_{2}=-  \mM_{22}^{-1} \mM_{21}\bc_{1}\,.
\ee

The reduced generalized eigenvalue problem \Eq{Eq:Diag} for the RSs can be further modified to the standard eigenvalue problem solved by matrix diagonalization, as it was done  in~\cite{MuljarovEPL10,DoostPRA14}. Calculating the effective perturbation matrix $\widetilde{\mM}_{11}$ requires inversion of a $N_2 \times N_2$ matrix $\mM_{22}$ with the compute time of about
$\Tinv=\Tinv_0(N_1 N_2)^{3/2}$, and we find using Matlab 2017a on 16 CPU cores (dual Intel E5-2640 v3) the value $\Tinv_0=7.5$\,ps. Solving \Eq{Eq:Diag} involves a matrix diagonalization which requires a compute time of about $\Tdia=\Tdia_0 N_1^3$, with $\Tdia_0=115$\,ps. For an equal number of RSs and static modes, the treatment of the static modes is thus about 15 times faster.

Now consider that the influence of static modes on a given $\varkappa$ is decreasing with increasing $\lambda$ only due to a reduction of the overlap matrix elements in $M$, owing to the increasing spatial frequency of the static RSs, see \Eq{Eq:Vnm}. For the RSs instead, there is additionally an increase of the wave number difference  $|k_n-\varkappa|$,
helping to reduce their influence with increasing $k_n$, see Eq.\,(38) of \Onlinecite{DoostPRA14}.
It is therefore beneficial for the accuracy of the RSE at a given compute time to use the lower numerical complexity of treating static modes to increase their number. We balance the numerical complexity of static modes and RSs in the RSE by choosing $\Tdia\approx\Tinv$, adjusting $N_1$ and $N_2$ appropriately.

To choose the basis RSs we use a cut-off in their spatial frequency inside the system, $|\sqrt{\epsilon}k_n|<\kmax$, with the maximum wave number $\kmax$ in vacuum, and the refractive index $\sqrt{\epsilon}$ of the basis sphere. The number of RSs is then approximately given by $N_1 \approx C_1^{(p)} \left(\sqrt{\epsilon} R\kmax\right)^d$, with the dimensionality $d\in\{1,2,3\}$, and the polarization $p\in\{{\rm TE,TM}\}$.  The values of $C_1^{(p)}$ determined for $\epsilon=4$ and $R\kmax\gg1$ are given in \Tab{tab:ADAS}.

\begin{table}
	\begin{tabular}{c | c | c | c}
		Dim. &  {\rm 1D} & {\rm 2D}  & {\rm 3D}\\
		\hline
		$C_1^{\rm TE}$	& 0.3125 & 0.1141 & 0.04258\\
		\hline
		$C_1^{\rm TM}$	& 0.325 & 0.1144 & 0.04287\\
		\hline
		VC  $C_2$ &	6.712 & 3.434 & 1.945\\
		\hline
		VSC $C_2$ &	6.196 & 3.276 & 1.857\\
	\end{tabular}
	\caption{Scaling parameters $C_1$  ($C_2$) of the number of basis RSs (static modes) for $\epsilon=4$ and different dimensionalities.}
	\label{tab:ADAS}
\end{table}

For the static modes, we use $\lambda<\kmaxS$, with a separate maximum wave number $\kmaxS$. The $\lambda=0$ static modes are always included in the VSC set. The resulting number of static modes is $N_2 \approx C_2 (R \kmaxS)^d$, with the values  of $C_2$ given in \Tab{tab:ADAS}, determined for $R\kmaxS \gg1$. We note that the scaling constants are not significantly different between VSC and VC sets. Requiring $\Tdia=\Tinv$, we then find
\be
\kmaxS=\left(\frac{\Tdia_0 C_1^{3/2}}{\Tinv_0 C_2^{3/2}}\right)^\frac{2}{3d} \sqrt{\epsilon}\kmax,
\ee
where $C_1=C_1^{\rm TM}$ for the 1D TM case, $C_1=C_1^{\rm TE}$ for the 1D TE case, and $C_1=C_1^{\rm TM}+C_1^{\rm TE}$ for the 2D and 3D case. We use this relation between $\kmaxS$ and $\kmax$ for all numerical results reported in the present work.


\begin{thebibliography}{1}

\bibitem{MuljarovEPL10}
E.~A. Muljarov, W. Langbein, and R. Zimmermann, Europhys. Lett. {\bf 92},
  50010  (2010).

\bibitem{LobanovPRA18}
S.~V. Lobanov, W. Langbein, and E.~A. Muljarov, Phys. Rev. A {\bf 98},  033820
  (2018).

\bibitem{DoostPRA14}
M.~B. Doost, W. Langbein, and E.~A. Muljarov, Phys. Rev. A {\bf 90},  013834
  (2014).

\bibitem{LaiPRA90}
H.~M. Lai, P.~T. Leung, K. Young, P.~W. Barber, and S.~C. Hill, Phys. Rev. A {\bf 41},  5187  (1990).

\bibitem{DoostPRA13}
M.~B. Doost, W. Langbein, and E.~A. Muljarov, Phys. Rev. A {\bf 87},  043827
  (2013).

\bibitem{Bohren1998}
C. Bohren and D.~R. Huffman, {\em Absorption and Scattering of Light by Small
  Particles} (Wiley Science Paperback Series, ADDRESS, 1998).

\bibitem{MuljarovOL18}
E.~A. Muljarov and T. Weiss, Opt. Lett. {\bf 43},  1978  (2018).

\bibitem{DoostPRA12}
M.~B. Doost, W. Langbein, and E.~A. Muljarov, Phys. Rev. A {\bf 85},  023835
  (2012).

\bibitem{MuljarovPRB16}
E.~A. Muljarov and W. Langbein, Phys. Rev. B {\bf 93},  075417  (2016).

\end{thebibliography}

\end{document}